\documentclass[a4paper,11pt]{article}
\usepackage{pos,amsmath}

\title{Influence of scattering versus
coherent parton branching on the
$k_T$ broadening of QCD cascades in a
medium
}
\author*[a]{M. Rohrmoser}
\author[a]{K. Kutak}
\author[b]{W. Płaczek}
\author[a]{E. Blanco}
\author[c]{R. Straka}

\affiliation[a]{Institute of Nuclear Physics, Polish Academy of Sciences,\\
ul. Radzikowskiego 152, 31-342 Krak\'{o}w, Poland}

\affiliation[b]{Institute of Applied Computer Science, Jagiellonian University,\\
ul.\ \L{}ojasiewicza 11, 30-348 Krak\'ow, Poland}
\affiliation[c]{AGH University of Science and Technology,\\ Krak\'ow, Poland}

\emailAdd{martin.rohrmoser@ifj.edu.pl}
\emailAdd{krzysztof.kutak@ifj.edu.pl}
\emailAdd{Wieslaw.Placzek@uj.edu.pl}
\emailAdd{eti.bla@orange.fr}
\emailAdd{strakrob@gmail.com}

\abstract{We studied the evolution of jets within a medium that contains both, transverse kicks
as well as medium induced coherent radiation.
In this framework parton branching occurs simultaneously to scatterings within the medium, leading to the interference effects that reproduce the well known BDMPS-Z emission rates and sizeable transverse momentum broadening.
We examined the relative importances of transverse momentum broadening from the coherent splittings and different types of in-medium scatterings and found a clear hierarchy of the influences from different scattering effects and deflections during branchings.}

\FullConference{%
  *** The European Physical Society Conference on High Energy Physics (EPS-HEP2021), ***\\
  *** 26-30 July 2021 ***\\
  *** Online conference, jointly organized by Universität Hamburg and the research center DESY ***
}

\begin{document}
\maketitle

%\section{...}
The hot and dense medium of a quark-gluon plasma (QGP) can be recreated in ultrarelativistic heavy ion collisions.
Since this medium cannot be accessed directly, due to the confinement of quarks and gluons, some more indirect methods need to be used to probe the medium.
One suitable type of probe are jets, highly energetic, collimated, strongly interacting sprays of particles.
The advantages of jets are that the initial jet particles are created at high energy densities, i.e. at early stages of the heavy ion collisions, interact with medium via processes of the strong interaction, yet do not thermalize, due to the high momentum and energy scales of the jet-particles involved.
Possible processes of jet medium interaction are scatterings of jet-particles off medium particles as well as processes of parton emission induced by scatterings off medium particles.
However, during the formation of a medium induced emission it is possible that the particles involved  undergo multiple scatterings off medium particles, giving rise to interference effects.
Spectra for this coherent medium induced radiations were first found (in the context of a QCD medium) by Baier, Dokshitzer, Mueller, Peign\'e, Schiff and independently by Zakharov (BDMPS-Z)
~\cite{Baier:2000mf,Baier:2000sb,Zakharov:1996fv,Zakharov:1997uu,Zakharov:1999zk,Baier:1994bd,Baier:1996vi}
.
 Later, an effective splitting kernel for coherent medium induced radiations for gluons as jet-particles was derived by Blaizot, Dominguez, Iancu, and Mehtar-Tani (BDIM)~\cite{Blaizot:2012fh}
as
\begin{equation}
{\cal K}(\mathbf{Q},z,p_0^+)=\frac{2}{p_0^+}\frac{P_{gg}(z)}{z(1-z)}\sin\left[\frac{\mathbf{Q}^2}{2k_{br}^2}\right]\exp\left[-\frac{\mathbf{Q}^2}{2k_{br}^2}\right] 
\label{eq:Kqz}
\end{equation}
with
\begin{equation}
\omega=xp_0^+,\,\,\,\, k_{br}^2=\sqrt{\omega_0\hat q_0},\,\,\,\,\,\mathbf{Q}=\mathbf{k}-z\,\mathbf{q},\,\,\, \omega_0=z(1-z)p_0^+ 
\end{equation}
and
\begin{equation}
\,\,\hat q_0=\hat q f(z),\,\, f(z)=1-z(1-z),\,\,\, P_{gg}(z)=N_c\frac{\left[1-z(1-z)\right]^2}{z(1-z)},    
\end{equation}
where $p_0^+=E$ is the energy of the initial jet particle, $x$ the parton momentum fraction (with regard to the initial energy $p_0^+$, $\mathbf{k}$ the jet-particle momentum components transverse to the jet axis, $\hat{q}$ the average transverse momentum transfer, $\alpha_S$ the QCD coupling constant and $N_C$ the number of colors.
Using the above splitting kernel, together with a scattering kernel $w$ (which will be defined further below) the following evolution equation over time $t$ can be derived for the fragmentation functions $D$ of jet-gluons in the medium~\cite{Blaizot:2013vha,Blaizot:2014rla}
\begin{equation}
\begin{aligned}
\frac{\partial}{\partial t} D(x,\mathbf{k},t) = & \:  \alpha_s \int_0^1 dz\, \int\frac{d^2q}{(2\pi)^2}\left[2{\cal K}(\mathbf{Q},z,\frac{x}{z}p_0^+) D\left(\frac{x}{z},\mathbf{q},t\right) 
- {\cal K}(\mathbf{q},z,xp_0^+)\, D(x,\mathbf{k},t) \right] \\
+& \int \frac{d^2\mathbf{l}}{(2\pi)^2} \,C(\mathbf{l})\, D(x,\mathbf{k}-\mathbf{l},t).
\end{aligned}
\label{eq:BDIM1}
\end{equation}
where
\begin{equation}
C(\mathbf{l}) = w(\mathbf{l}) - \delta(\mathbf{l}) \int d^2\mathbf{l'}\,w(\mathbf{l'})\,,
\label{eq:Cq}
\end{equation}
with the scattering kernels~\cite{Blaizot:2013vha,Blaizot:2014rla}
\begin{equation}
 w(\mathbf{l}) = \frac{16\pi^2\alpha_s^2N_cn}{\mathbf{l}^4}\,,
\label{eq:wq1}
\end{equation}
where $n$ is the density of scatterers in the medium
and  \cite{Aurenche:2002pd}
\begin{equation}
 w(\mathbf{l}) = \frac{g^2m_D^2T}{\mathbf{l}^2(\mathbf{l}^2+m_D^2)}\,,
\label{eq:wq2}
\end{equation}
with the Debye mass
$m_D^2=g^2T^2\left(\frac{N_c}{3}+\frac{N_f}{6}\right),\quad
$and $
g^2 = 4\pi\alpha_s.
$

With the Sudakov-factors
\begin{equation}
    \Delta(p_0^+,t)=\exp{\left(-t\left[\int_{|\mathbf{q}|>q_\downarrow} \frac{d^2\mathbf{q}}{(2\pi)^2}\left( w(\mathbf{q})+\alpha_s\int_0^{1-\epsilon} dz 2z{\cal K}(\mathbf{q},z,p_0^+)\right)\right]\right)}\,,
    \label{eq:sud}
\end{equation}
where the notation $|\mathbf{q}|>q_\downarrow$ should indicate that the integration runs over all $\mathbf{q}$ except those where $|\mathbf{q}|<q_\downarrow$,
the above integro-differential evolution equation, Eq.~(\ref{eq:BDIM1}), can be formulated as the following integral equation

\begin{align}
    D(x,\mathbf{k},t)&=D(x,\mathbf{k},t_0)\frac{\Delta(xp_0^+,t)}{\Delta(xp_0^+,t_0)}
    \nonumber\\
    &+\int_{t_0}^{t} dt'\frac{\Delta(xp_0^+,t)}{\Delta(xp_0^+,t')}\int_{|\mathbf{q}|>q_\downarrow} \frac{d^2\mathbf{q}}{(2\pi)^2}\int_0^{1-\epsilon} dz \int \frac{d^2\mathbf{Q}}{(2\pi)^2}\int_0^1 dy (2\pi)^2 
    \nonumber\\&
        \left[w(\mathbf{Q})\delta^{(2)}(\mathbf{k}-(\mathbf{Q}+\mathbf{q}) )\delta(x-y)+\alpha_s2z{\cal K}(\mathbf{Q},z,yp_0^+)\delta^{(2)}(\mathbf{k}-(\mathbf{Q}+z\mathbf{q}) )\delta(x-zy)\right]
        \nonumber\\&
        D(y,\mathbf{q},t')\,,
    \label{eq:int_eq_bdim}
\end{align}
in the simultaneous limits of $\epsilon\rightarrow 0$ and $q_\downarrow\rightarrow 0$.
The formulation of the evolution equation as an integral equation allows numerical solution of Eq.~ (\ref{eq:BDIM1})
by a Monte-Carlo algorithm~\cite{Kutak:2018dim,Blanco:2020uzy}
. A goal of the presented work~\cite{Blanco:2020uzy}
was to study the influences of non-collinear branchings and different types of scatterings in Eq.~(\ref{eq:BDIM1})
on the broadening of the distribution of transverse momentum $k_T=||\mathbf{k}||$.
To this end note that a collinear splitting kernel can be found as 
    \begin{equation}
       {\cal K}(z)=\int d^2\mathbf{Q}  {\cal K}(\mathbf{Q},z,yp_0^+)\frac{\sqrt{yp_0^+}}{2\pi\sqrt{\hat{q}}}=\frac{f(z)^{5/2}}{(z(1-z))^{3/2}}\,.
       \label{eq:fromKzQtoKz}
    \end{equation}
and a corresponding evolution equation~\cite{Blaizot:2013vha,Blaizot:2014rla}
can also be formulated as
    \begin{equation}
\begin{aligned}
\frac{\partial}{\partial t} D(x,\mathbf{k},t) = & \: \frac{1}{t^*} \int_0^1 dz\, {\cal K}(z) \left[\frac{1}{z^2}\sqrt{\frac{z}{x}}\, D\left(\frac{x}{z},\frac{\mathbf{k}}{z},t\right)\theta(z-x) 
- \frac{z}{\sqrt{x}}\, D(x,\mathbf{k},t) \right] \\
+& \int \frac{d^2\mathbf{q}}{(2\pi)^2} \,C(\mathbf{q})\, D(x,\mathbf{k}-\mathbf{q},t),
\end{aligned}
\label{eq:BDIM2}
\end{equation}
 where 
    \begin{equation}
\frac{1}{t^\ast}=\frac{\alpha_s N_c}{\pi}\sqrt{\frac{\hat{q}}{p_0^+}}\,.
    \end{equation}
Integration over the transverse momenta yields the following evolution equation
    \begin{equation}
\begin{aligned}
\frac{\partial}{\partial t} D(x,t) = & \: \frac{1}{t^*} \int_0^1 dz\, {\cal K}(z) \left[\sqrt{\frac{z}{x}}\, D\left(\frac{x}{z},t\right)\theta(z-x) 
- \frac{z}{\sqrt{x}}\, D(x,t) \right]\,,
\end{aligned}
\label{eq:BDIM_coll}
\end{equation}
Thus, in order to study deviations from a possible gaussian broadening in transverse momenta, we can also construct the following case, where the momentum fractions $x$ follow a fragmentation function $D(x,t)$, that is described by Eq.~(\ref{eq:BDIM_coll})
while the transverse momenta $k_T$ are selected from a Gaussian distribution
so that the fragmentation function $D(x,k_T,t)$ is given by 
\begin{equation}
D(x,\mathbf{k},t) = D(x,t)\,\frac{4\pi}{\langle k_\perp^2\rangle}
\exp\left[-\frac{\mathbf{k}^2}{\langle k_\perp^2\rangle}\right],
\end{equation}
where 
\begin{equation}
\langle k_\perp^2\rangle=\min\left\{\frac{1}{2}\hat q t(1+x^2),\, \frac{k^2_{br}(x)}{4\bar\alpha},
\,(x E)^2\right\}, \quad k_{br}^2(x)=\sqrt{x E \hat q}.   
\end{equation}
In the above, it is assumed that $k_\perp^2<\omega^2=(xE)^2$.
We will refer to this case as Gaussian approximation.
The different cases of jet-medium interactions studied are
\begin{itemize}
\item the Gaussian approximation,
\item the collinear branching case $\mathcal{K}(z)$ following Eq.~(\ref{eq:BDIM2})
with scatterings given by Eq.~(\ref{eq:wq1}),
\item the collinear branching case $\mathcal{K}(z)$ following Eq.~(\ref{eq:BDIM2})
with scatterings given by Eq.~(\ref{eq:wq2}),
\item the non-collinear branching case $\mathcal{K}(z,\mathbf{Q})$ following Eq.~(\ref{eq:BDIM1})
without scatterings,
\item the non-collinear branching case $\mathcal{K}(z,\mathbf{Q})$ following Eq.~(\ref{eq:BDIM1})
with scatterings given by Eq.~(\ref{eq:wq1}),
\item the non-collinear branching case $\mathcal{K}(z,\mathbf{Q})$ following Eq.~(\ref{eq:BDIM1})
with scatterings given by Eq.~(\ref{eq:wq2}).
\end{itemize}  
For the numerical studies a constant coupling constant of $\alpha_s=\frac{\pi}{10}$ was assumed and 
the following parameters for a medium that is invariant in time
\begin{align}
\hat{q}&=1 \textrm{GeV}^2/\textrm{fm}\,,&&
n= 0.243\textrm{GeV}^3\,,\nonumber\\
m_D&= 0.993\textrm{GeV}\,,&&
p_0^+=100\textrm{GeV}\,.
\end{align}

\begin{figure}[!ht]
\centering{}
\includegraphics[width=0.32\textwidth]{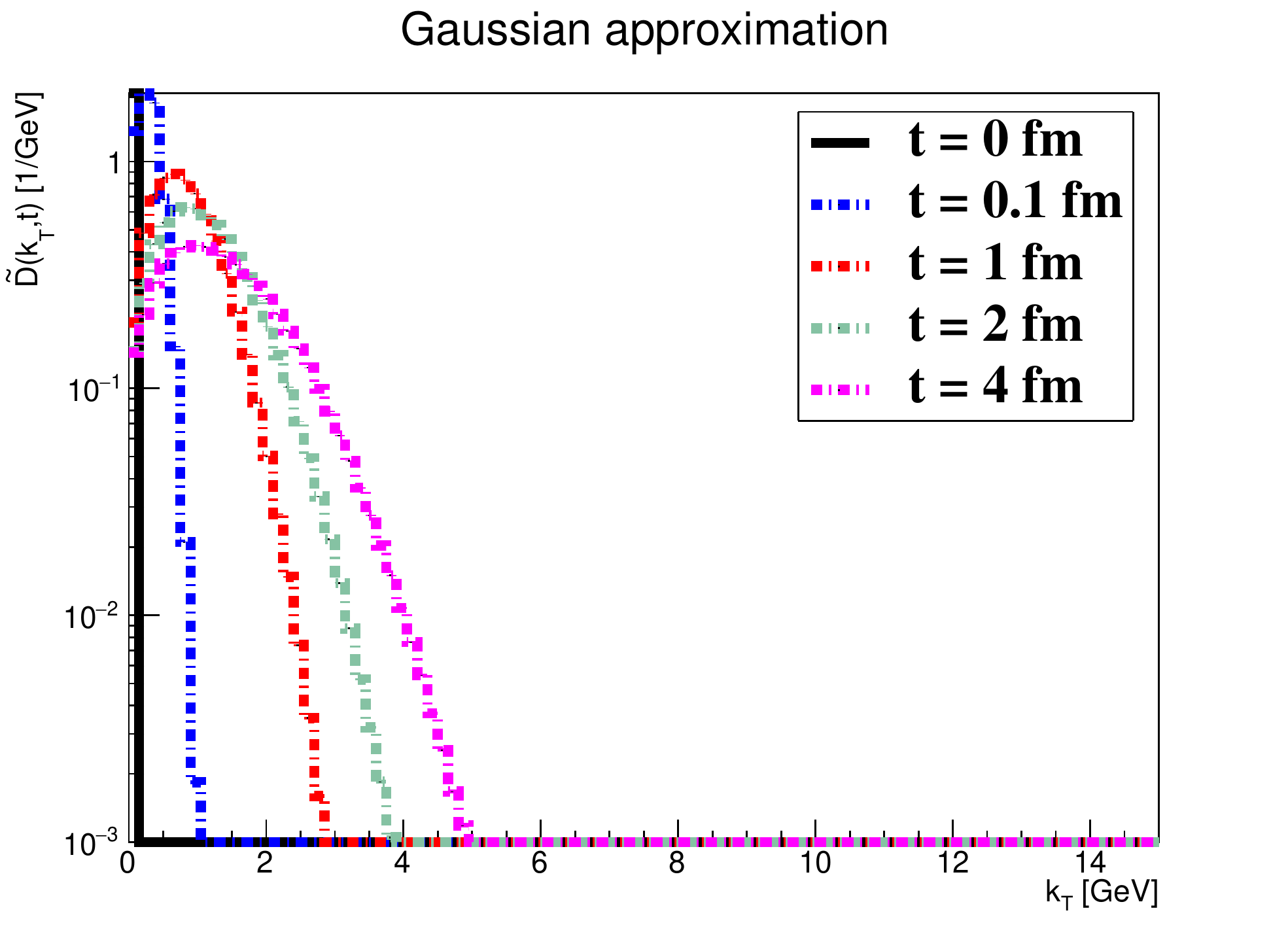}
\includegraphics[width=0.32\textwidth]{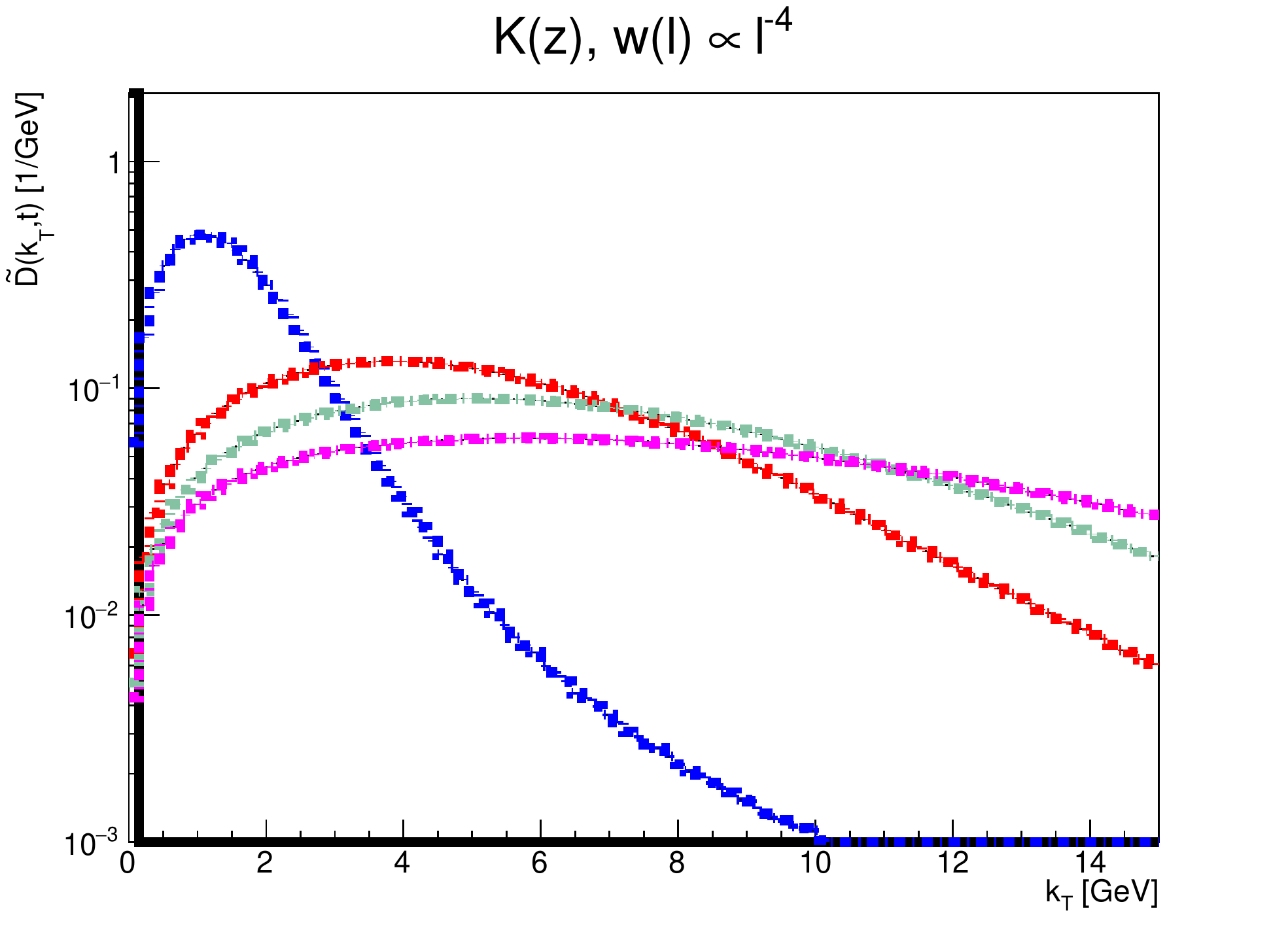}
\includegraphics[width=0.32\textwidth]{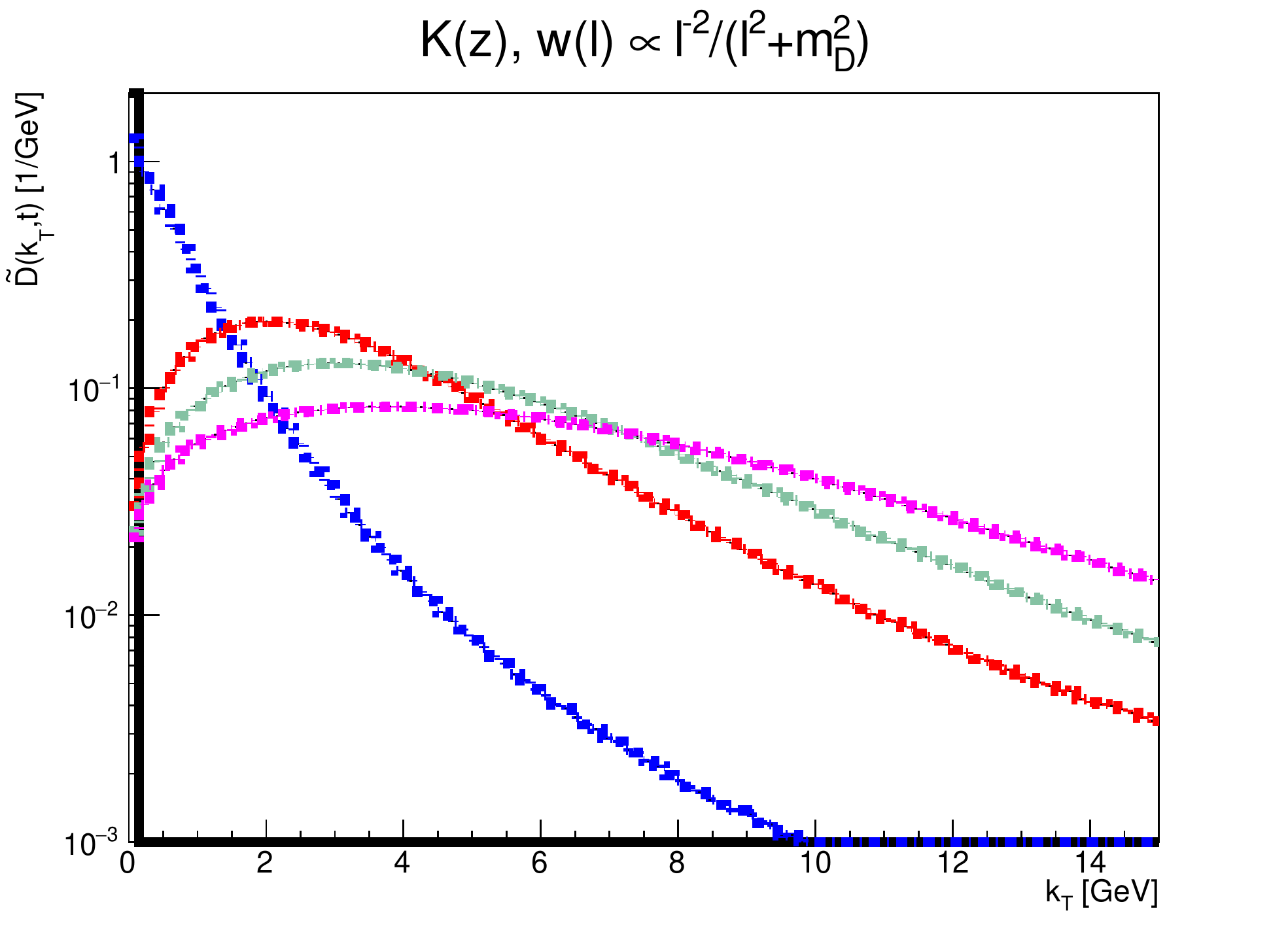}
\includegraphics[width=0.32\textwidth]{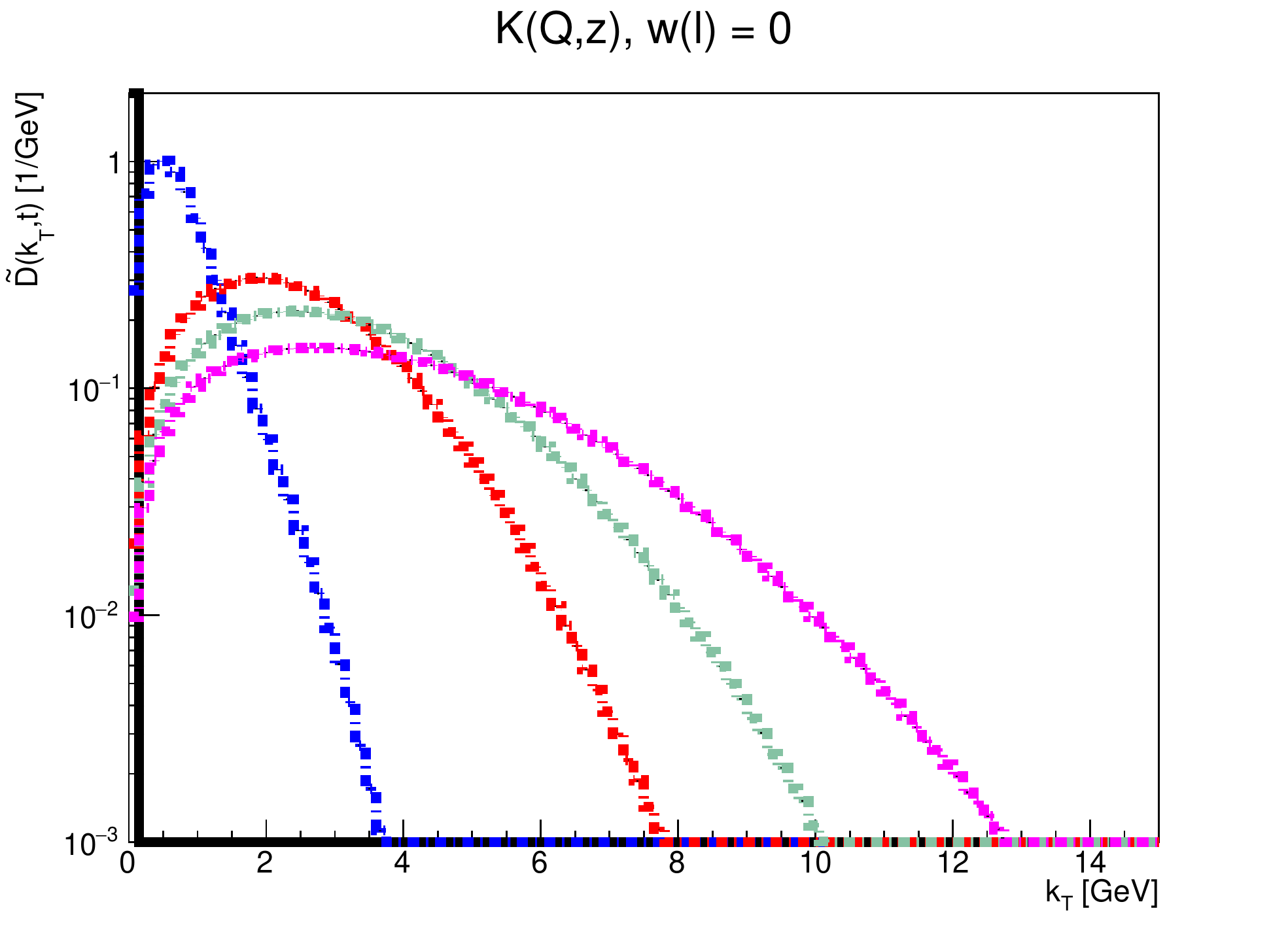}
\includegraphics[width=0.32\textwidth]{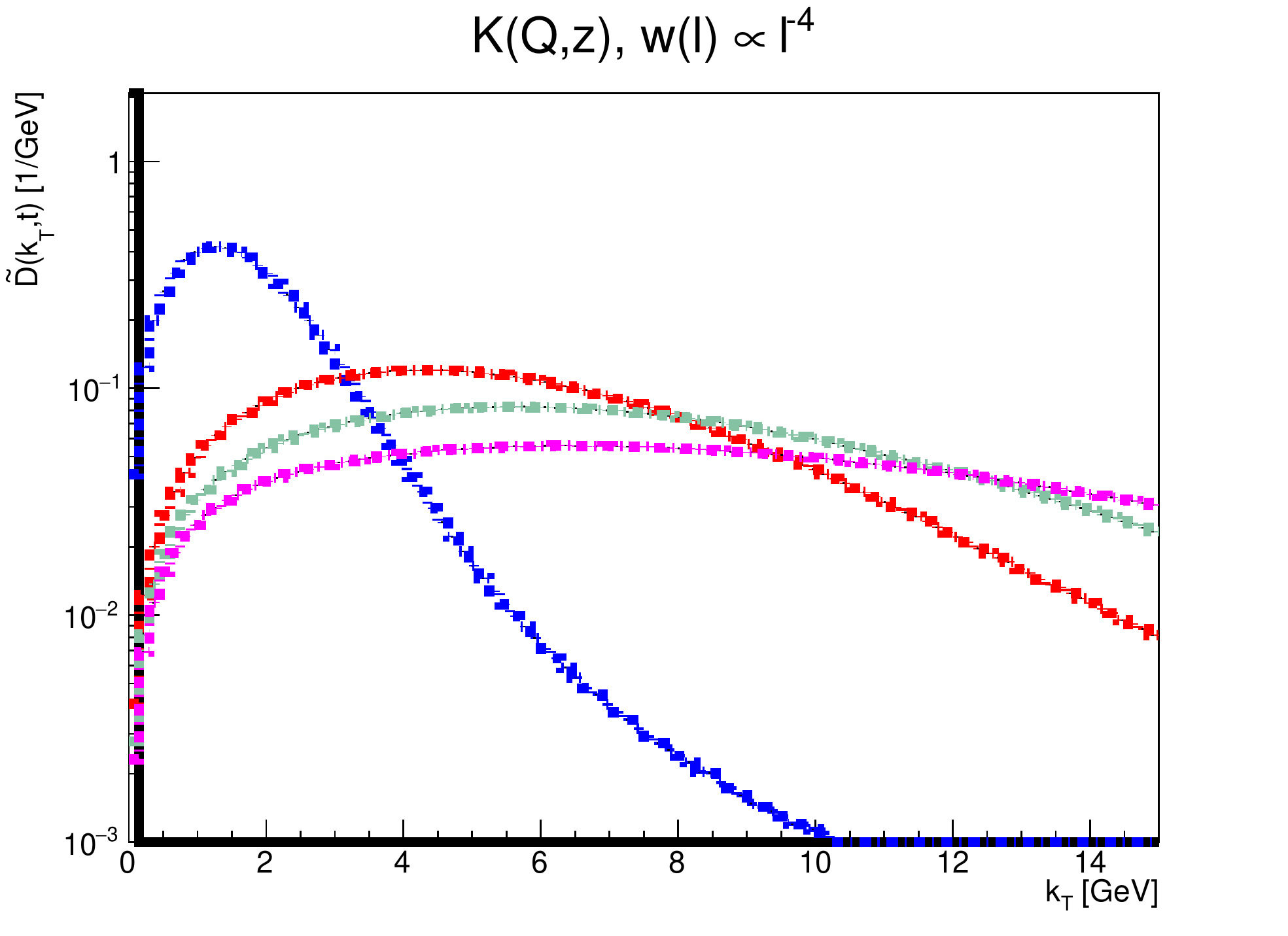}
\includegraphics[width=0.32\textwidth]{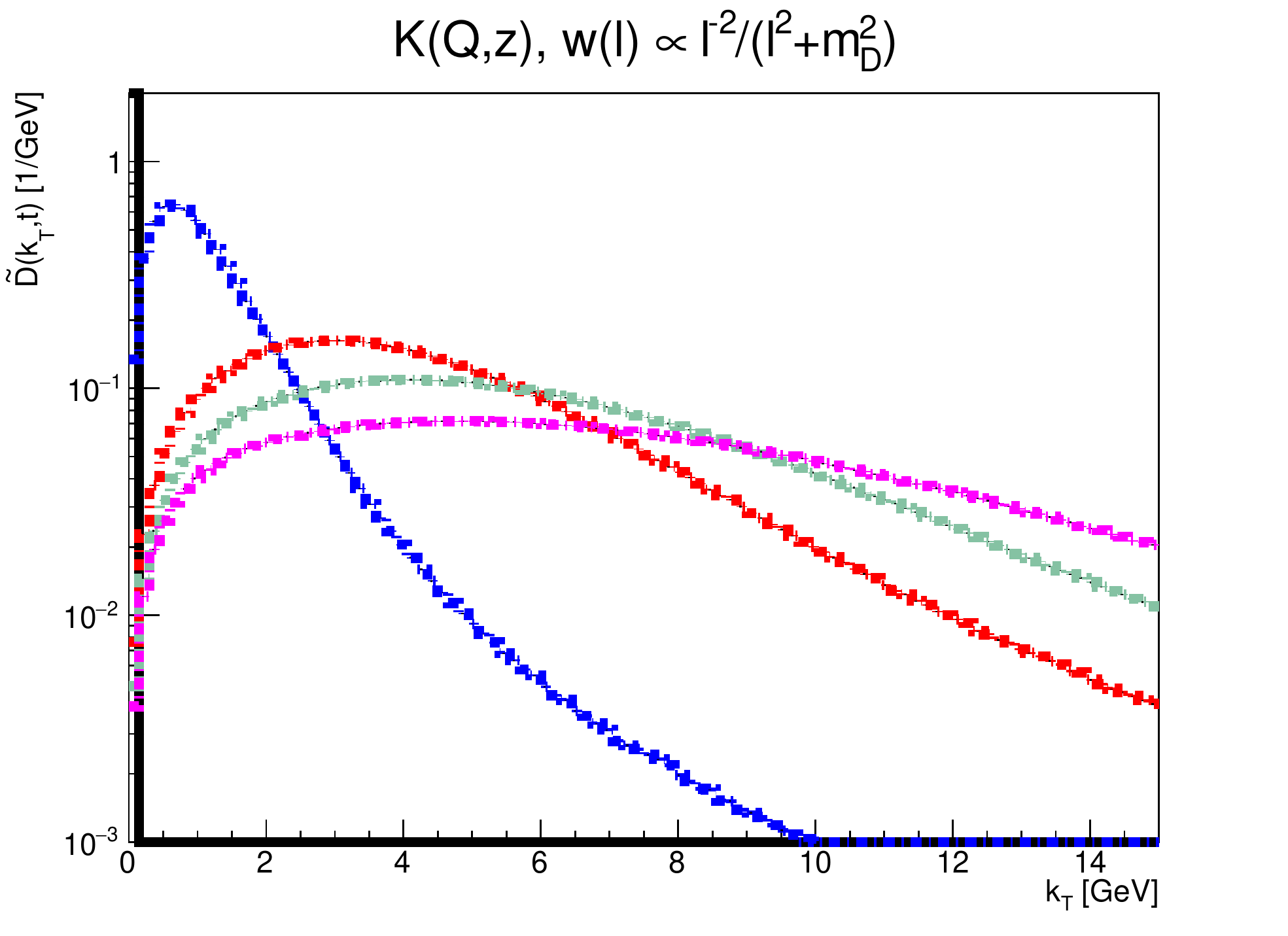}
\caption{The $k_T$ distributions for the evolution time values $t=0, 0.1, 1, 2, 4\,$fm, for different kernels: the Gaussian approximation, ${\cal K}(z)$ and  ${\cal K}(\mathbf{Q},z,p^+)$ (denoted as $\rm K(z)$ and $\rm K(Q,z)$, respectively), and different collision terms: no collision term, the collision term as in Eq.~(\ref{eq:wq1}) and as in Eq.~(\ref{eq:wq2}).}
\label{Fig1}
\end{figure}

\begin{figure}[!ht]
\centering{}
\includegraphics[width=0.32\textwidth]{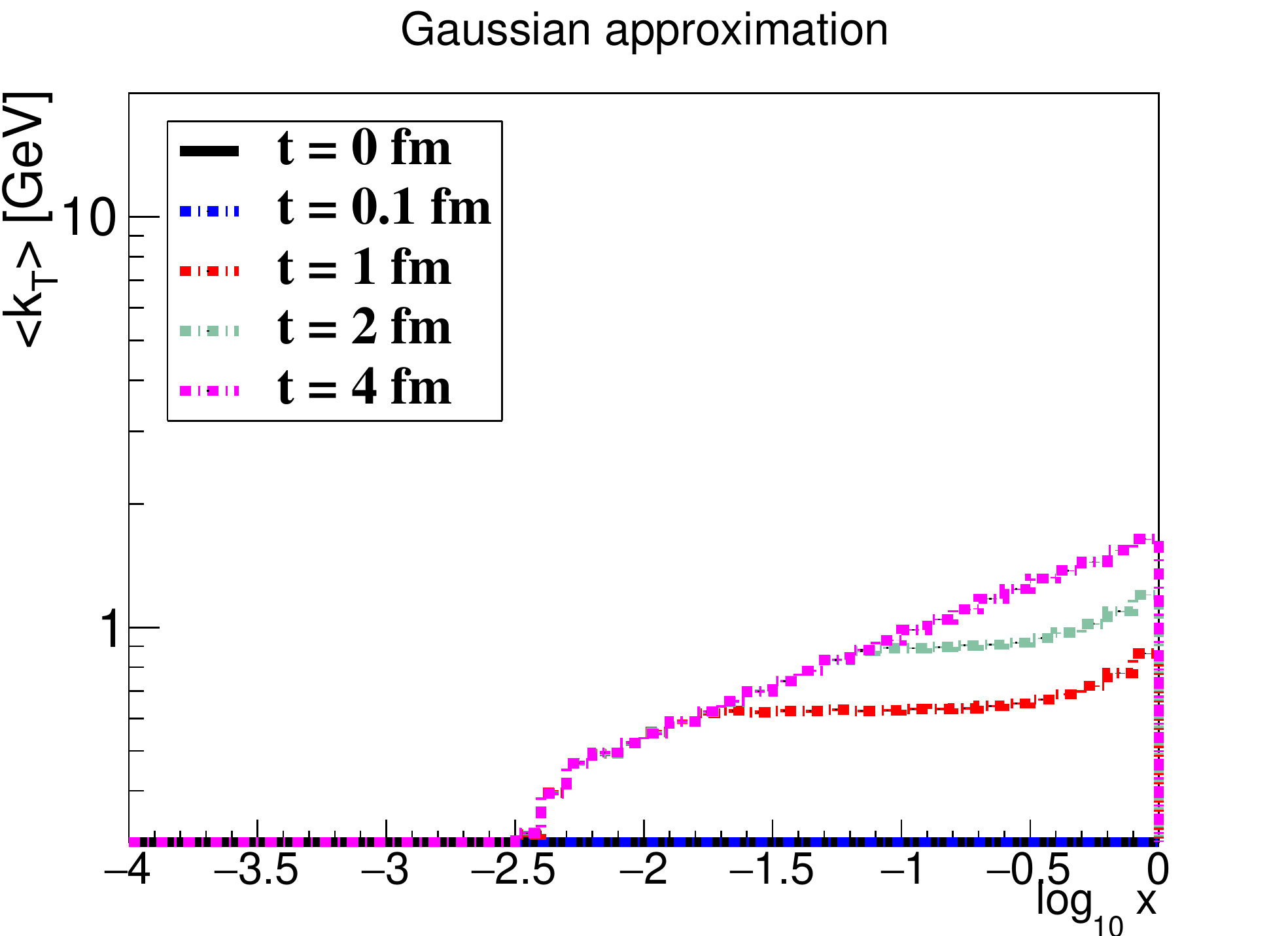}
\includegraphics[width=0.32\textwidth]{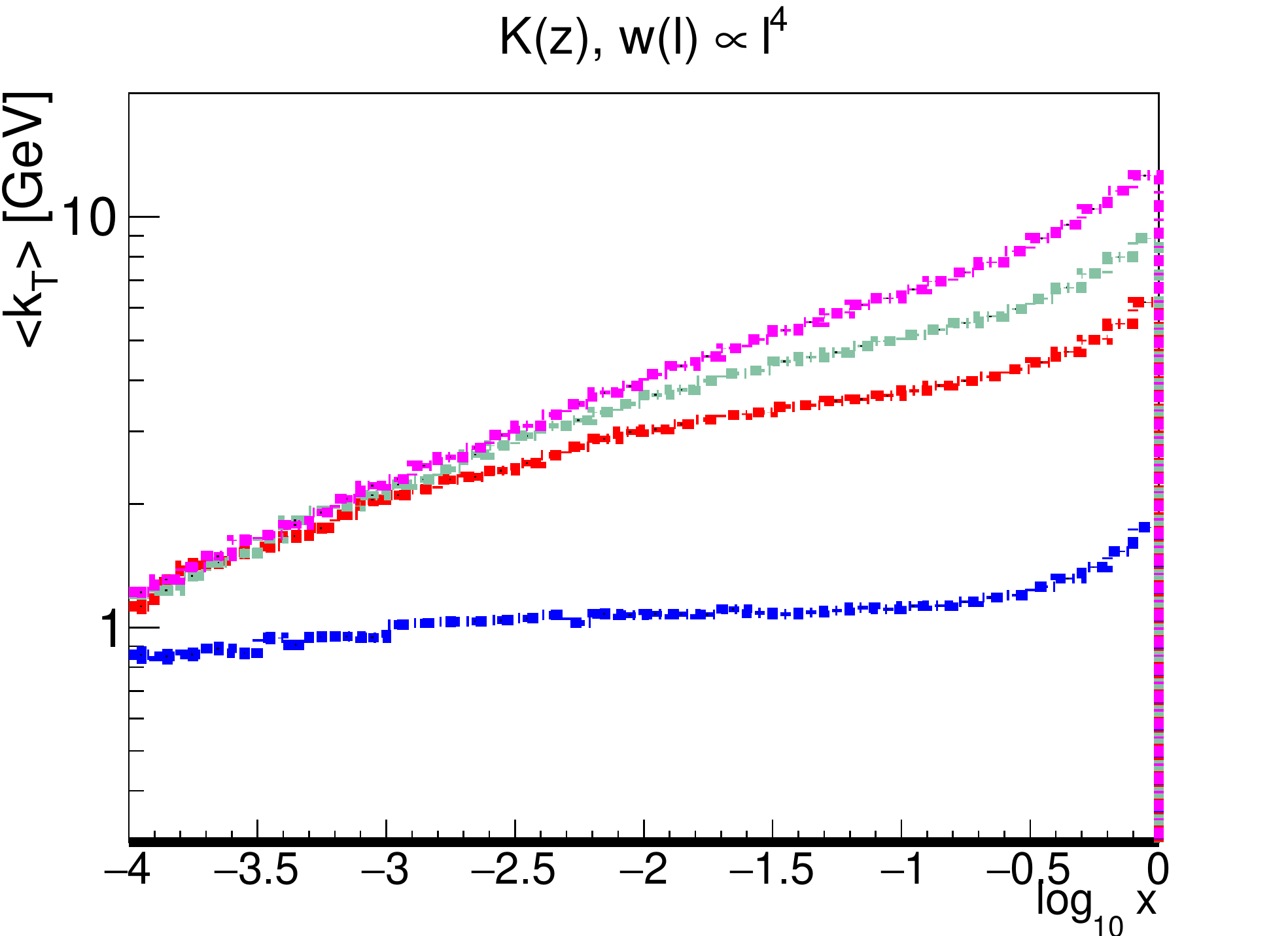}
\includegraphics[width=0.32\textwidth]{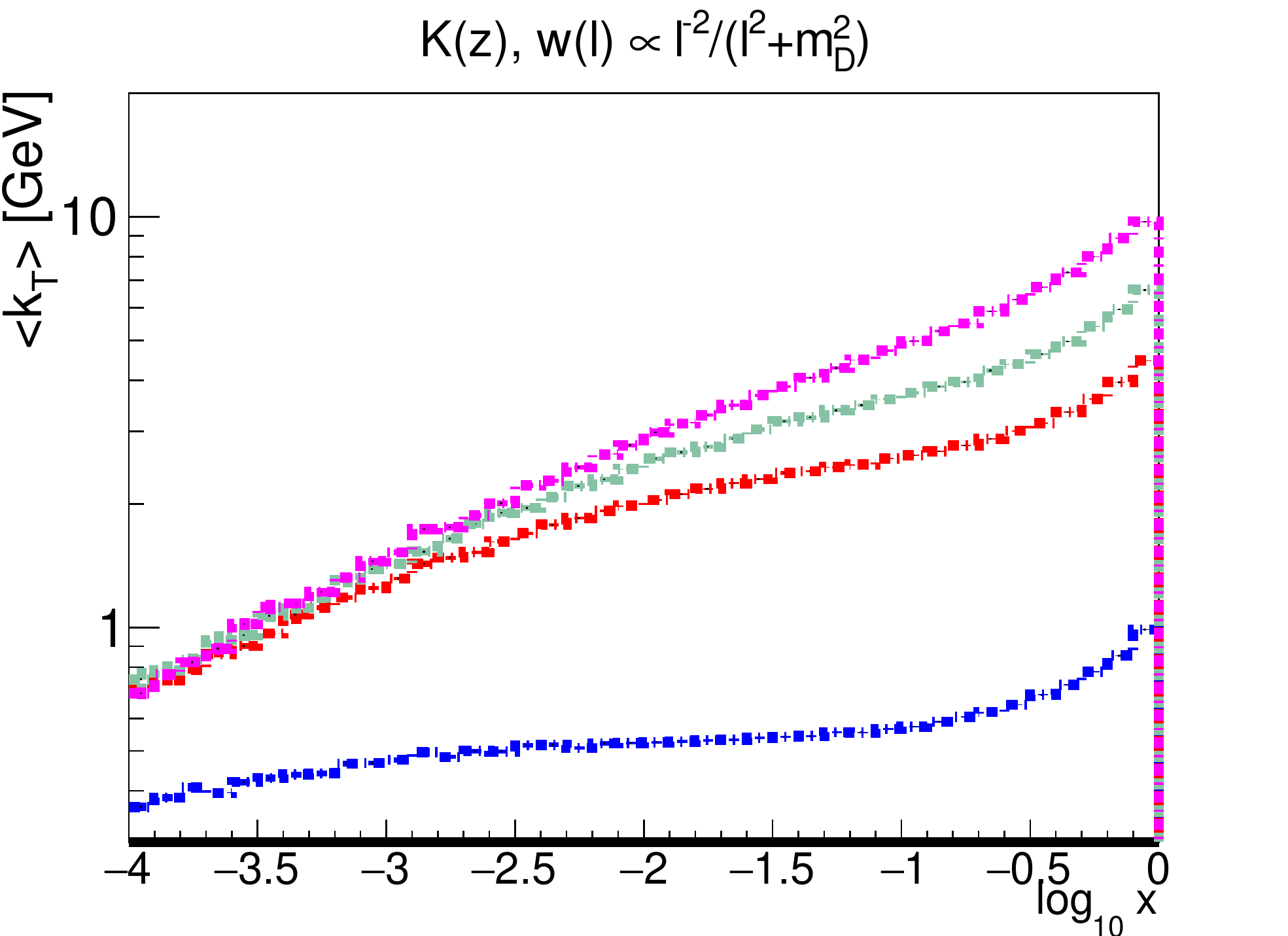}
\includegraphics[width=0.32\textwidth]{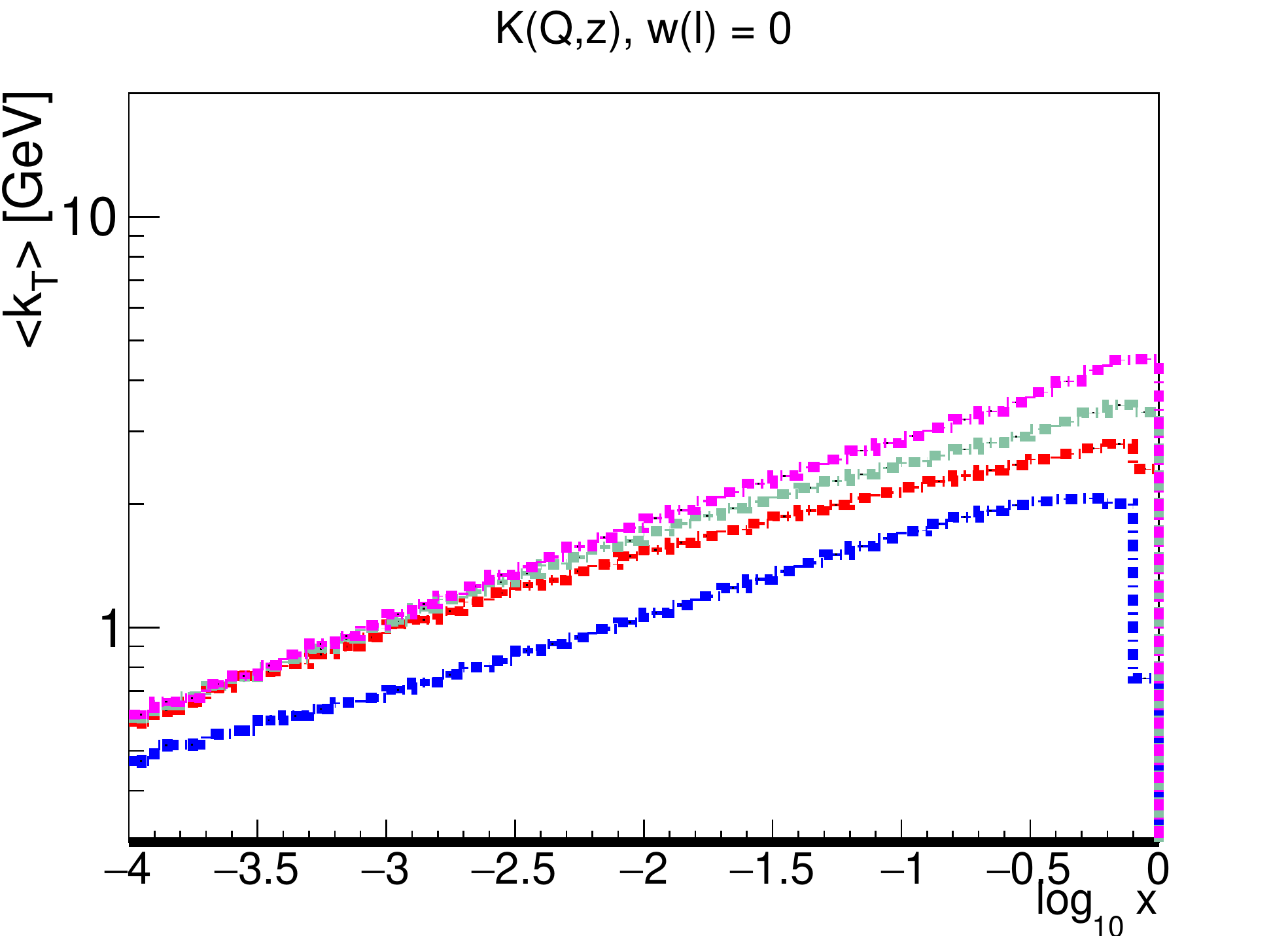}
\includegraphics[width=0.32\textwidth]{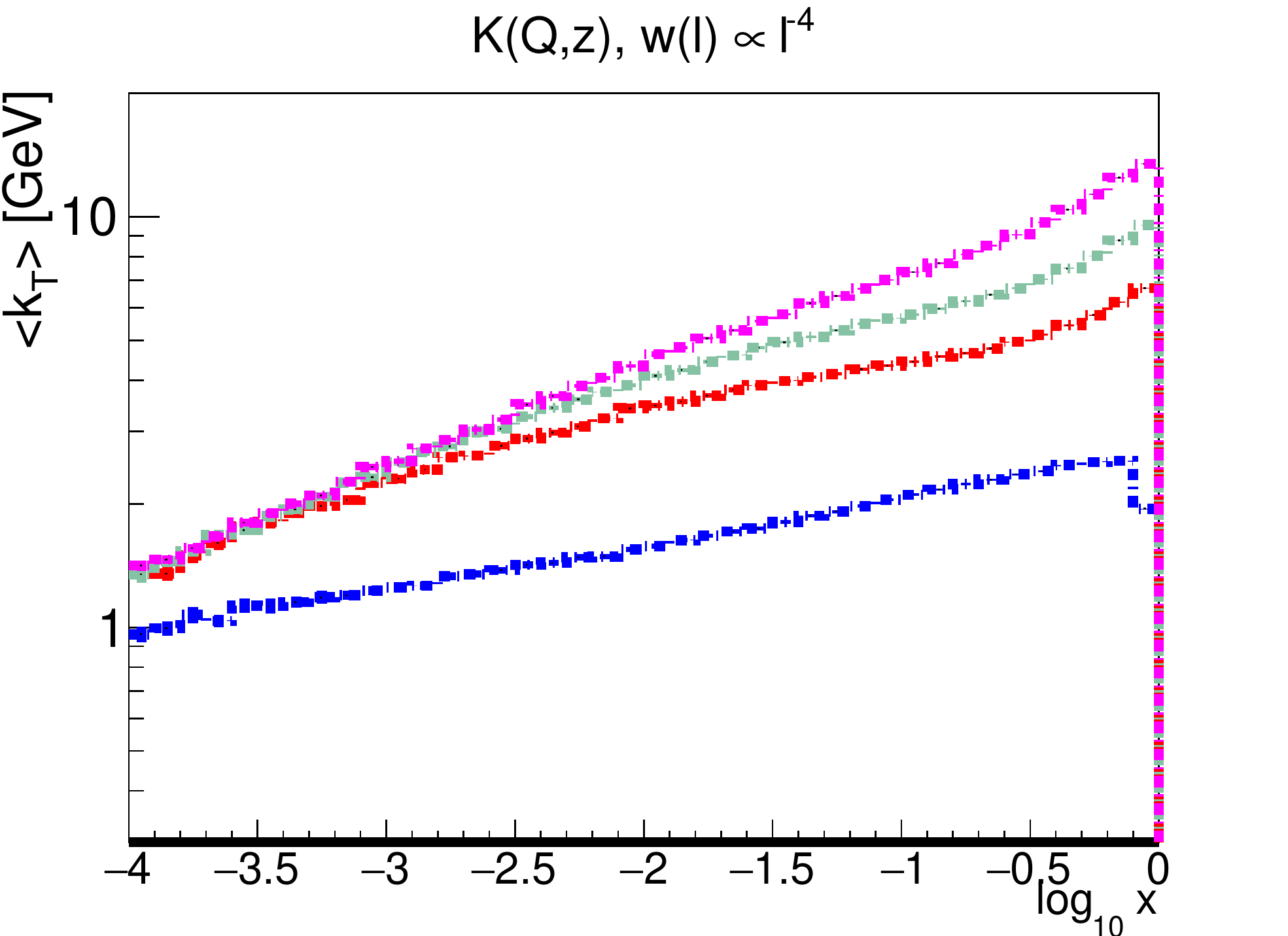}
\includegraphics[width=0.32\textwidth]{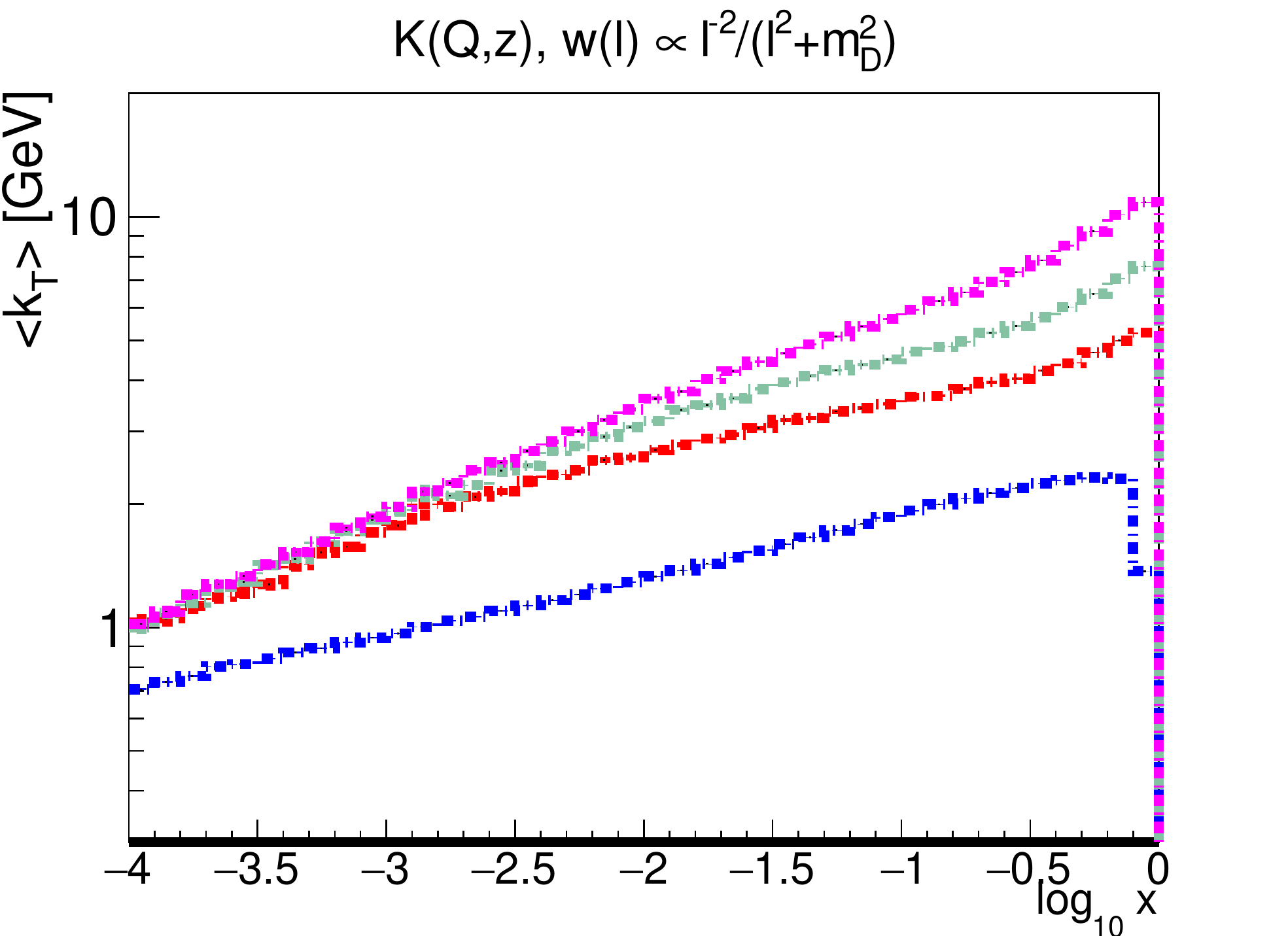}
\caption{The $\langle k_T \rangle$ vs.\ $\log_{10}x$ distributions for the evolution time values $t=0, 0.1, 1, 2, 4\,$fm, for different kernels: the Gaussian approximation, ${\cal K}(z)$ and  ${\cal K}(\mathbf{Q},z,p^+)$ (denoted as $\rm K(z)$ and $\rm K(Q,z)$, respectively), and different collision terms: no collision term, the collision term as in Eq. ~(\ref{eq:wq1}) and in Eq.~(\ref{eq:wq2}), respectively.}
\label{Fig2}
\end{figure}

Fig.~\ref{Fig1} shows results for the distributions
\begin{equation}
\Tilde{D}(x,k_T,t) = \int_0^{2\pi} %dk_\perp 
    d \phi\,k_T\,D(x,\mathbf{k},t)\,,\quad \textrm{with } k_T=||\mathbf{k}||\,.
\end{equation}
for different time-scales.
It can be seen that the distributions exhibit different broadenings in $k_T$ over time obeying the following ordering from smallest to largest broadening:
\begin{enumerate}
\item the Gaussian approximation,
\item the non-collinear branching case $\mathcal{K}(z,\mathbf{Q})$ following Eq.~(\ref{eq:BDIM1})
without scatterings,
\item the collinear branching case $\mathcal{K}(z)$ following Eq.~(\ref{eq:BDIM2})
with scatterings given by  Eq.~(\ref{eq:wq2}),
\item the non-collinear branching case $\mathcal{K}(z,\mathbf{Q})$ following Eq.~(\ref{eq:BDIM1})
with scatterings given byEq.~(\ref{eq:wq2}),
\item the collinear branching case $\mathcal{K}(z)$ following Eq.~(\ref{eq:BDIM2})
with scatterings given by Eq.~(\ref{eq:wq1}),
\item the non-collinear branching case $\mathcal{K}(z,\mathbf{Q})$ following Eq.~(\ref{eq:BDIM1})
with scatterings given by Eq.~(\ref{eq:wq1}).
\end{enumerate}
Furthermore, the $k_T$ broadening can be also studied via its average value $\langle k_T\rangle$ for jet-particles with a given momentum fraction $x$.
Results are shown in Fig.~\ref{Fig2}.
Again, the different types of jet-medium interactions yield different broadenings in $k_T$ over time, following the same ordering as before.
As can be seen at low energy scales a common behavior can be found for large time-scales, corresponding to a approximately polynomial rise in $\langle k_T\rangle$.

\section{Summary}
We have studied the in-medium evolution of gluon jets vie processes of coherent medium induced radiations as well as scatterings off medium particles.
To this end, numerical solutions for the fragmentation functions that follow the BDIM-evolution equations~Eqs.~(\ref{eq:BDIM1}),~(\ref{eq:BDIM2}), and~(\ref{eq:BDIM_coll})~\cite{Blaizot:2013vha,Blaizot:2014rla}, 
were obtained. 
It was found that the resulting distributions for transverse momenta $k_T$ exhibit a clear ordering of their broadening with regard to the influences of non-collinear branchings and scatterings off medium particles: The dominant influences on broadening come from scattering effects. Non-collinear branchings yield a non-negligible, however smaller broadening effect.
\acknowledgments
This work was partially supported by the Polish National Science Centre with the grant no.\ DEC-2017/27/B/ST2/01985.
\bibliographystyle{JHEP}
\bibliography{refs1}

\providecommand{\href}[2]{#2}\begingroup\raggedright\begin{thebibliography}{10}

\bibitem{Baier:2000mf}
R.~Baier, D.~Schiff and B.G.~Zakharov, \emph{{Energy loss in perturbative
  QCD}}, \href{https://doi.org/10.1146/annurev.nucl.50.1.37}{\emph{Ann. Rev.
  Nucl. Part. Sci.} {\bfseries 50} (2000) 37}
  [\href{https://arxiv.org/abs/hep-ph/0002198}{{\ttfamily hep-ph/0002198}}].

\bibitem{Baier:2000sb}
R.~Baier, A.H.~Mueller, D.~Schiff and D.T.~Son, \emph{{'Bottom up'
  thermalization in heavy ion collisions}},
  \href{https://doi.org/10.1016/S0370-2693(01)00191-5}{\emph{Phys. Lett. B}
  {\bfseries 502} (2001) 51}
  [\href{https://arxiv.org/abs/hep-ph/0009237}{{\ttfamily hep-ph/0009237}}].

\bibitem{Zakharov:1996fv}
B.G.~Zakharov, \emph{{Fully quantum treatment of the Landau-Pomeranchuk-Migdal
  effect in QED and QCD}}, \href{https://doi.org/10.1134/1.567126}{\emph{JETP
  Lett.} {\bfseries 63} (1996) 952}
  [\href{https://arxiv.org/abs/hep-ph/9607440}{{\ttfamily hep-ph/9607440}}].

\bibitem{Zakharov:1997uu}
B.G.~Zakharov, \emph{{Radiative energy loss of high-energy quarks in finite
  size nuclear matter and quark - gluon plasma}},
  \href{https://doi.org/10.1134/1.567389}{\emph{JETP Lett.} {\bfseries 65}
  (1997) 615} [\href{https://arxiv.org/abs/hep-ph/9704255}{{\ttfamily
  hep-ph/9704255}}].

\bibitem{Zakharov:1999zk}
B.G.~Zakharov, \emph{{Transverse spectra of radiation processes in-medium}},
  \href{https://doi.org/10.1134/1.568149}{\emph{JETP Lett.} {\bfseries 70}
  (1999) 176} [\href{https://arxiv.org/abs/hep-ph/9906536}{{\ttfamily
  hep-ph/9906536}}].

\bibitem{Baier:1994bd}
R.~Baier, Y.L.~Dokshitzer, S.~Peigne and D.~Schiff, \emph{{Induced gluon
  radiation in a QCD medium}},
  \href{https://doi.org/10.1016/0370-2693(94)01617-L}{\emph{Phys. Lett. B}
  {\bfseries 345} (1995) 277}
  [\href{https://arxiv.org/abs/hep-ph/9411409}{{\ttfamily hep-ph/9411409}}].

\bibitem{Baier:1996vi}
R.~Baier, Y.L.~Dokshitzer, A.H.~Mueller, S.~Peigne and D.~Schiff, \emph{{The
  Landau-Pomeranchuk-Migdal effect in QED}},
  \href{https://doi.org/10.1016/0550-3213(96)00426-9}{\emph{Nucl. Phys. B}
  {\bfseries 478} (1996) 577}
  [\href{https://arxiv.org/abs/hep-ph/9604327}{{\ttfamily hep-ph/9604327}}].

\bibitem{Blaizot:2012fh}
J.-P.~Blaizot, F.~Dominguez, E.~Iancu and Y.~Mehtar-Tani, \emph{{Medium-induced
  gluon branching}}, \href{https://doi.org/10.1007/JHEP01(2013)143}{\emph{JHEP}
  {\bfseries 01} (2013) 143} [\href{https://arxiv.org/abs/1209.4585}{{\ttfamily
  1209.4585}}].

\bibitem{Blaizot:2013vha}
J.-P.~Blaizot, F.~Dominguez, E.~Iancu and Y.~Mehtar-Tani, \emph{{Probabilistic
  picture for medium-induced jet evolution}},
  \href{https://doi.org/10.1007/JHEP06(2014)075}{\emph{JHEP} {\bfseries 06}
  (2014) 075} [\href{https://arxiv.org/abs/1311.5823}{{\ttfamily 1311.5823}}].

\bibitem{Blaizot:2014rla}
J.-P.~Blaizot, L.~Fister and Y.~Mehtar-Tani, \emph{{Angular distribution of
  medium-induced QCD cascades}},
  \href{https://doi.org/10.1016/j.nuclphysa.2015.03.014}{\emph{Nucl. Phys.}
  {\bfseries A940} (2015) 67}
  [\href{https://arxiv.org/abs/1409.6202}{{\ttfamily 1409.6202}}].

\bibitem{Aurenche:2002pd}
P.~Aurenche, F.~Gelis and H.~Zaraket, \emph{{A Simple sum rule for the thermal
  gluon spectral function and applications}},
  \href{https://doi.org/10.1088/1126-6708/2002/05/043}{\emph{JHEP} {\bfseries
  05} (2002) 043} [\href{https://arxiv.org/abs/hep-ph/0204146}{{\ttfamily
  hep-ph/0204146}}].

\bibitem{Kutak:2018dim}
K.~Kutak, W.~P\l{}aczek and R.~Straka, \emph{{Solutions of evolution equations
  for medium-induced QCD cascades}},
  \href{https://doi.org/10.1140/epjc/s10052-019-6838-9}{\emph{Eur. Phys. J. C}
  {\bfseries 79} (2019) 317}
  [\href{https://arxiv.org/abs/1811.06390}{{\ttfamily 1811.06390}}].

\bibitem{Blanco:2020uzy}
E.~Blanco, K.~Kutak, W.~P\l{}aczek, M.~Rohrmoser and R.~Straka, \emph{{Medium
  induced QCD cascades: broadening and rescattering during branching}},
  \href{https://doi.org/10.1007/JHEP04(2021)014}{\emph{JHEP} {\bfseries 04}
  (2021) 014} [\href{https://arxiv.org/abs/2009.03876}{{\ttfamily
  2009.03876}}].

\end{thebibliography}\endgroup

\end{document}